\documentclass{ptapap}
\usepackage{amsmath}
\usepackage{amssymb}
\usepackage{color}
\usepackage{xcolor}
\usepackage{hyperref}
\usepackage{subcaption}
\usepackage{graphicx}

\def\hb{{\sc{H}}$\beta$\/}
\def\mgii{Mg{\sc{ii}}}

\author{Swayamtrupta Panda}[CFT, CAMK]
\author{Bo\.zena Czerny}[CFT]
\affil[CAMK]{Nicolaus Copernicus Astronomical Center, Polish Academy of Sciences, Bartycka 18, 00--716 Warsaw, Poland}
\affil[CFT]{Center for Theoretical Physics, Polish Academy of Sciences, Al. Lotnik\'ow 32/46, 02--668 Warsaw, Poland}

\title{Generating AGN spectra for LSST : First steps}

\begin{document}

\maketitle

\begin{abstract}

In the near future the arrival of new surveys promises a significant increment in the number of quasars (QSO) at large redshift ranges. This will help to constrain the dark energy models using quasars. The Large Synoptic Survey Telescope (\textit{LSST}) will cover over 10 million QSO in six photometric bands during its 10-year run. QSO will be monitored and subsequently analysed using the photometric reverberation mapping technique. Oncoming \textit{LSST} data will decrease the uncertainties in the dark energy determination using reverberating--mapped sources, particularly at high redshifts. We present the first steps in modelling of light curves for \hb\ and \mgii\ and discuss the quasar selection in the context of photometric reverberation mapping with \textit{LSST}. With the onset of the \textit{LSST} era, we expect a huge rise in the overall quasar counts and redshift range covered ($z\leq7.0$), which will provide a better constraint of AGN properties with cosmological purposes.

\end{abstract}

\section{Photo-reverberation mapping with \textit{LSST}}

\begin{figure}
\begin{subfigure}{.5\textwidth}
  \centering
  \includegraphics[width=1.3\linewidth]{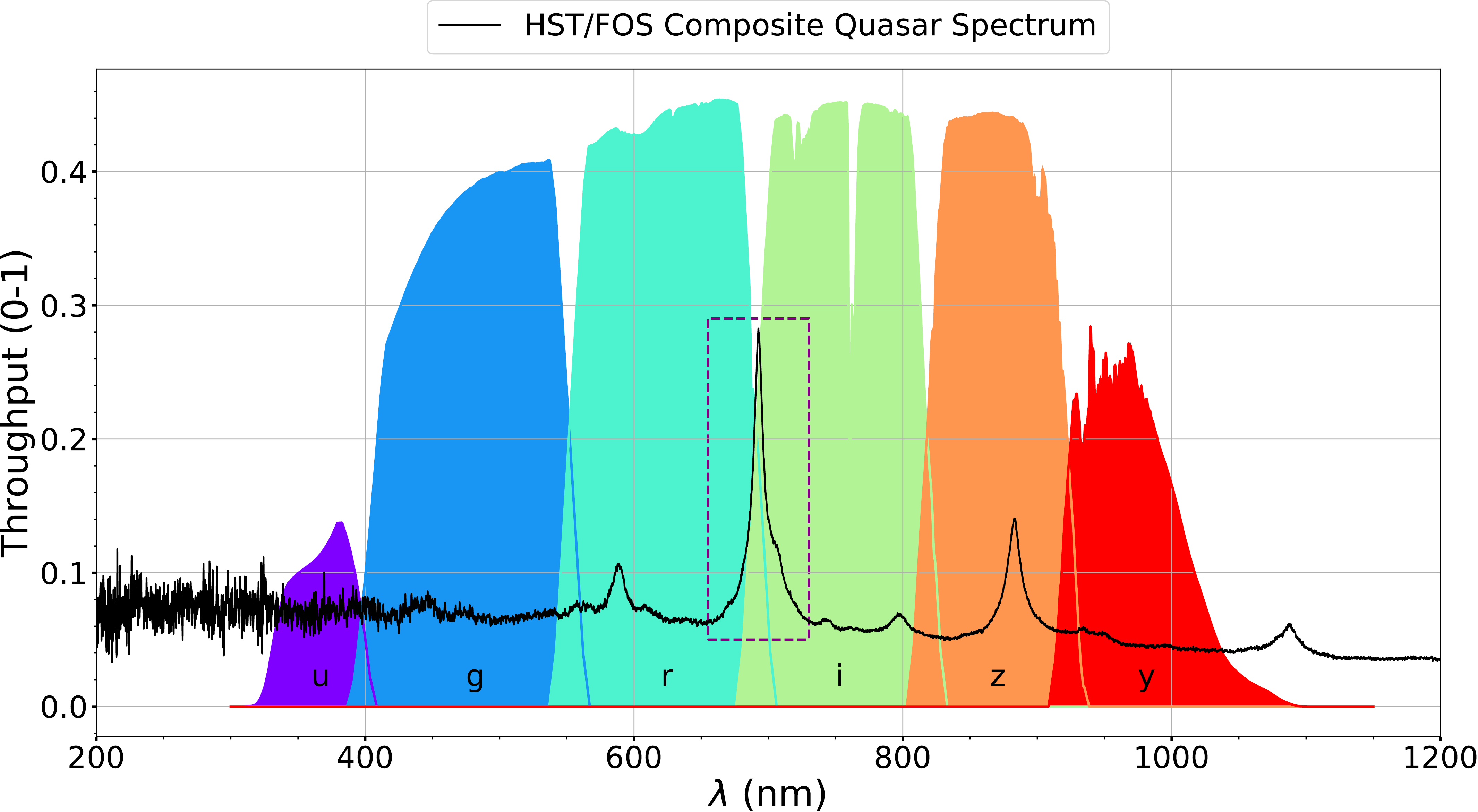}
\end{subfigure}%
\quad\quad\quad
\begin{subfigure}{.5\textwidth}
  \centering
  \includegraphics[width=.625\linewidth]{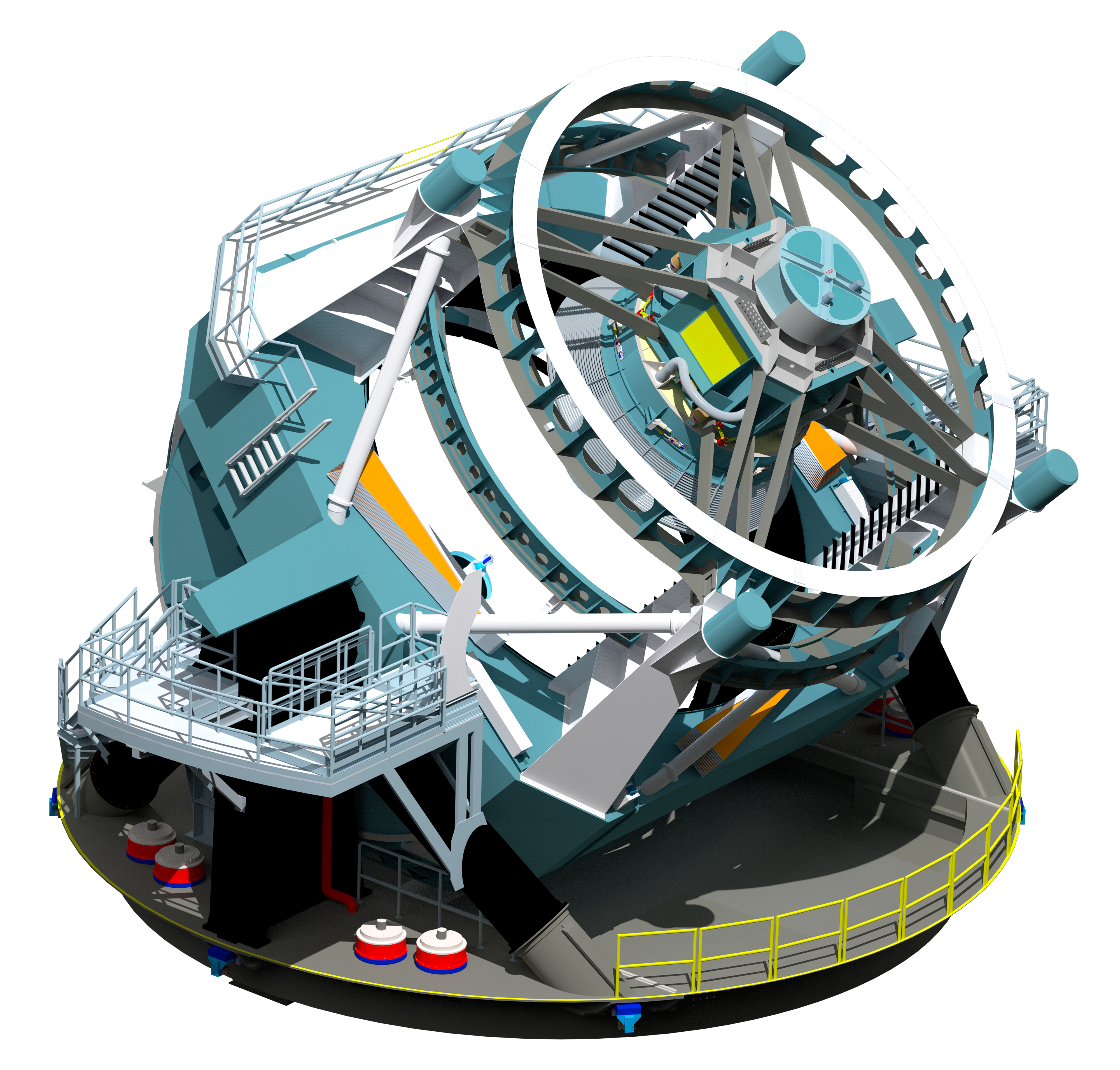}
\end{subfigure}
\caption{LEFT: Throughput curves for the 6 \textit{LSST} photometric bands (\textit{ugrizy}), shown as a function of wavelength (in \AA). The black solid line is the \textit{HST/FOS} composite quasar spectrum and the dashed box in purple highlights the importance of quasar selection. The spectrum is shown at a redshift, \textit{z} = 2.56. The composite spectrum is downloaded from \href{https://archive.stsci.edu/prepds/composite_quasar/}{https://archive.stsci.edu/prepds/composite\_quasar/}. RIGHT: A three dimensional rendering of the baseline design for the \textit{LSST} with the telescope pointed at an elevation of about 45 degrees.}
\label{fig:fig}
\end{figure}

Reverberation mapping is a robust observational technique to determine the time-lag $\tau_0$ between the variability of the continuum emission of an active galactic nucleus (AGN) and the line emission associated with the broad-line region \citep[BLR, ][]{1982ApJ...255..419B, 1997ASSL..218...85N}. \textit{LSST} will be a public optical/NIR survey of $\sim$half the sky in the \textit{ugrizy} bands (see Figure \ref{fig:fig}) up to r$\sim$27.5 based on $\sim820$ visits across all the six photometric bands over a 10-year period \citep{lsst2019}. The 8.4m telescope with a state-of-the-art 3.2 Gigapixel flat-focal array camera will allow to perform rapid scan of the sky with 15 seconds exposure and thus providing a moving array of color images of objects that change. The whole observable sky is planned to be scanned every $\sim$4 nights. It is expected that upon the completion of the main-survey period, \textit{LSST} will have mapped $\sim$20 billion galaxies and $\sim$17 billion stars using these six photometric bands. Every night, \textit{LSST} will monitor $\sim$75 million AGNs and is estimated to detect $\sim$300+ million AGNs in the $\sim$18000 deg$^2$ main-survey area.

The \textit{LSST} AGN survey will produce a high-purity sample of at least 10 million well-defined, optically-selected AGNs. Utilizing the large sky coverage, depth, the six filters extending to 1$\rm{\mu}$m, and the valuable temporal information of \textit{LSST}, this AGN survey will supersede the largest current AGN samples by more than an order of magnitude. Each region of the \textit{LSST} sky will receive $\sim$200 visits in each band in the decade long monitoring, allowing variability to be explored on timescales from minutes to a decade (see Chapter 10 of \citet{lsst_book} for more details). \textit{LSST} will conduct more intense observations of at least 4 \textit{Deep Drilling Fields} or DDFs, i.e. COSMOS, XMM-LSS, W-CDF-S and ELAIS-S1, each of which cover $\sim$10 deg$^2$ (more details in \citealt{lsst_wp}). \citet{lsst_wp} estimates an increase in the depth in the DDFs by over an order of magnitude relative to the main-survey. This will be made possible owing to a $\sim$18-20 times increase in the number of visits per photometric band.

The proposed cadence ($\sim$2 nights) of the observations of AGNs and the expected re-visits during the 10-year run will allow to (1) ensure a high lag recovery fraction for the relatively short accretion-disk lags expected \citep{Yu2018}; and (2) allow investigations of secular evolution of these lags that test the underlying model. \textit{LSST} will also perform quality accretion disk reverberation mapping for $\sim$3000 AGNs in the DDFs.

\section{Suitability of quasar}
\label{sub:suitability}
In order to increase the purity of the sample in the \textit{LSST}, especially in the DDFs, we would want to constrain the quasar counts in terms of only Type 1 AGNs with respect to the observed broad emission lines such as H$\beta$, MgII, CIV. We perform a simple filtering to estimate these numbers in terms of total predicted number of quasars to be observed over the full duration of \textit{LSST}. This filtering is based on three quantities that affect our estimation of the ``good'' Type 1 quasars, namely (1) the line intensity; (2) overlap of the photometric bands; and (3) the line full-width at half maximum (FWHM). The peak intensity of the line (including continuum) that will be procured from the pipeline is already folded with the filters and to recover the true intensity one needs to scale by $\frac{1}{T}$, where \textit{T} is the throughput (or transmittance) of the corresponding band which is estimated from the band function. We estimate that $\sim$26\% (for MgII) and $\sim$36\% (for H$\beta$) of the supposed Type 1 quasars will end up being observed in these overlapping windows taking into account the respective emission lines broadening and our naive assumption of the throughput cutoff \citep[see][for more details]{2019arXiv190905572P}. 

\section{Modelling `real' light curves}
\label{sub:modelling}
\textit{LSST} is a photometric project but the 6-channel photometry (see Figure \ref{fig:fig}) can be effectively used for the purpose of reverberation mapping and estimation of time delays. We are developing a code that takes into consideration several key parameters to produce AGN light curves in the context of \textit{LSST}, namely -- (1) the campaign duration of the instrument (10 years); (2) number of visits per photometric band; (3) the photometric accuracy (0.01-0.1 mag); (4) black hole mass distribution; (5) luminosity distribution; (6) redshift distribution; and (7) line equivalent widths (EWs) consistent with SDSS quasar catalogue \citep{s11}. Each predicted photometric magnitude is calculated taking into account the emission from the lines (H$\beta$, MgII, CIV) as well as the in the continuum. The code also incorporates the FeII pseudo-continuum and contamination from starlight i.e. stellar contribution \citep[see][for more details]{2019arXiv190905572P, 2019arXiv191002725L}. 

\section{Future}
\label{sub:future}
The code in development, although at a basic stage, is able to produce decent estimates to time-delay measurements. We are currently working to extend the analyses to include all six photometric channels simultaneously, and retrieve the time delay measurements to mimic actual observations. Finally, we plan to create lightcurves incorporating a realistic distribution of the redshift, luminosity, $M_{BH}$ and EWs. This will be useful -- first for preparing a mock catalog of quasar lightcurves, and second, to test real data that will be obtained from the \textit{LSST} in the near future.

\acknowledgements{The project was partially supported by National Science Centre, Poland, grant No. 2017/26/A/ST9/00756 (Maestro 9), and by the Ministry of Science and Higher Education (MNiSW) grant DIR/WK/2018/12.}

\bibliographystyle{ptapap}
\bibliography{panda2}

\end{document}